\documentclass[letterpaper]{article}
\usepackage[T1]{fontenc}

\usepackage{geometry}
\usepackage{setspace}

\usepackage[articletitle=true]{achemso}

\usepackage{graphicx}
\usepackage{float}
\newfloat{scheme}{htbp}{los}
\floatname{scheme}{Scheme}
\floatname{chart}{Chart}
\newfloat{graph}{htbp}{loh}

\usepackage{chemformula} 
\usepackage[version = 4]{mhchem} 

\usepackage{authblk}
\author[1]{Giacomo Mariani*}
\author[1]{Riccardo Lodo}
\author[1]{Keigo Matsuyama}
\author[1]{Yoji Kunihashi}
\author[1]{Taro Wakamura}
\author[1]{Satoshi Sasaki}
\author[2]{Louis Smet}
\author[2,3,4]{Makoto Kohda}
\author[1,2]{Junsaku Nitta}
\author[1]{Haruki Sanada}

\affil[1]{NTT Basic Research Laboratories, NTT, Inc., Atsugi 243-0198, Japan}
\affil[2]{Department of Materials Science, Graduate School of Engineering, Tohoku University, Sendai 980-8579, Japan}
\affil[3]{Center for Science and Innovation in Spintronics, Tohoku University, Sendai 980-8577, Japan}
\affil[4]{Quantum Materials and Applications Research Center, National Institutes for Quantum Science and Technology, Takasaki 370-1292, Japan}

\title{Uniform Narrow Excitonic Spectrum in Large-Area Suspended WSe$_2$ Monolayers}
\date{*Email: giacomo.mariani@ntt.com}

\usepackage{pdfpages}

\begin{document}

\maketitle

\begin{abstract}
Uniformity in the excitonic spectrum is a key requirement for accessing intrinsic excitonic physics in two-dimensional semiconductors; however, in transition-metal dichalcogenide (TMD) monolayers supported on substrates, exciton energies and linewidths can vary spatially due to inhomogeneities from contact with other materials or fabrication residues. Suspended TMD monolayers provide a route to minimizing substrate-induced disorder, although conventional transfer processes can introduce contamination. Here we demonstrate the spatially uniform excitonic spectrum from optically high-quality WSe$_{2}$ suspended monolayers fabricated by gold-assisted exfoliation directly onto an Au contact electrode of a gate-tunable device. The resulting membranes span narrow suspended regions up to $\sim$80 $\mu$m and show spatially uniform photoluminescence at cryogenic temperatures with neutral-exciton linewidths as low as $\sim$4.5 meV. Spectral reproducibility supports an intrinsic optical response, while gate-dependent measurements resolve multiple excitonic species. This approach provides a route to electrically tunable potential landscapes in suspended TMD monolayers with a highly uniform excitonic response.
\end{abstract}
\\

Excitons in two-dimensional transition-metal dichalcogenide (TMD) monolayers (MLs) have attracted intense interest for both fundamental studies and optoelectronic applications \cite{mak2010atomically, splendiani2010emerging, chernikov2014exciton, wang2018colloquium}. Spectrally narrow excitonic resonances are essential for selectively addressing individual optical transitions, enabling controlled manipulation by external electric and magnetic fields, and isolating the contribution of specific excitonic complexes in spatio-temporal dynamics studies \cite{cadiz2017excitonic, robert2017fine, courtade2017charged}. However, even with narrow linewidths, spatial inhomogeneities in the potential landscape (arising from strain and dielectric disorder) can obscure intrinsic behavior, complicate quantitative optical studies, and limit device scalability. 

Achieving both narrow excitonic linewidths and spatially uniform optical response over extended areas remains a central challenge in TMD monolayers. A widely used route to improve optical quality is hexagonal boron-nitride (hBN) encapsulation, which screens disorder and suppresses inhomogeneous broadening. hBN-encapsulated WSe$_{2}$ and related TMD MLs can exhibit neutral-exciton linewidths as narrow as 3–5 meV at cryogenic temperatures \cite{cadiz2017excitonic, wierzbowski2017direct, ajayi2017approaching}. However, van der Waals assembly often traps residues that form interfacial nanobubbles or blisters, producing localized strain potentials and spatially inhomogeneous excitonic emission \cite{khestanova2016universal,raja2019dielectric,darlington2020imaging}. As a result, excitonic properties that are locally pristine may vary significantly across a single heterostructure, highlighting the importance of achieving exciton uniformity over length scales relevant for spatially resolved optical spectroscopy and device applications. 

Suspended TMD MLs offer a conceptually attractive route to achieving spatially uniform excitonic response by eliminating substrate-induced disorder and dielectric screening. In principle, removing the supporting substrate should promote homogeneous optical properties across extended areas. In practice, however, suspended devices, both with and without gate control, have primarily relied on polymer transfer over etched holes or trenches \cite{morell2016high, hernandez2022strain, kumar2024strain, onodera2024drytransfer, kumar2025strain, wu2025modulation}. Such approaches can introduce contamination and interfacial residues that broaden excitonic resonances and hinder systematic investigations of exciton uniformity in suspended membranes.

To address these limitations, we focus on a transfer-free gold-assisted exfoliation (GAE) approach to fabricate suspended WSe$_{2}$ MLs directly onto pre-patterned Au contact electrodes in a gate-tunable device geometry. GAE was first introduced as a one-step polymer-free method to realize large-area MLs on gold surfaces, exploiting the strong interaction between bulk TMDs and freshly prepared smooth Au films \cite{magda2015exfoliation, desai2016gold, velicky2018mechanism, huang2020universal, ziewer2025strain}. Subsequent studies extended this approach to other metallic films \cite{fu2022one, sun2022exfoliation, grubivsic2023situ, han2025hybridization}. GAE has also been applied to directly exfoliate suspended MLs on trenches or cavities \cite{rodriguez2022activation, huang2022efficient, geilen2025situ}; however, exciton properties in suspended regions have been primarily evaluated via PL measurements, where both room-temperature \cite{huang2022efficient} and low-temperature \cite{wang2022intensive} studies report only broad linewidths on the order of tens of meV. Such measurements are not sufficient for a quantitative assessment of local disorder and potential fluctuations. Moreover, previous implementations do not incorporate electrostatic control, which is essential to separate different excitonic complexes and remove ambiguities arising from spectral overlap.

Here we investigate the spatial uniformity of excitonic emission in gate-tunable suspended WSe$_2$ MLs at cryogenic temperature, enabled by a direct and transfer-free GAE technique. We employ GAE to avoid the common contamination pathways associated with conventional transfer techniques, while remaining compatible with electrostatic gating. Using cryogenic, spatially resolved PL, exciton uniformity is quantified through statistical analysis of exciton energies and linewidths. We assess the reproducibility of these quantities across extended suspended regions and across spatially separated areas extending over millimeter-scale distances, as well as the influence of electrostatic gating on the excitonic spectra in a spatially uniform system. The suspended MLs span holes and long trenches extending up to $\sim$80 $\mu$m and exhibit neutral-exciton linewidths of $\sim$4.5 meV, comparable to the narrow linewidths reported for high-quality hBN-encapsulated MLs. The resulting electrically tunable potential landscape with high spatial uniformity over large areas provides a promising platform for studies of long-range exciton transport and device applications in TMD monolayers.

\begin{figure*}
  \centering
  \includegraphics[width=\linewidth]{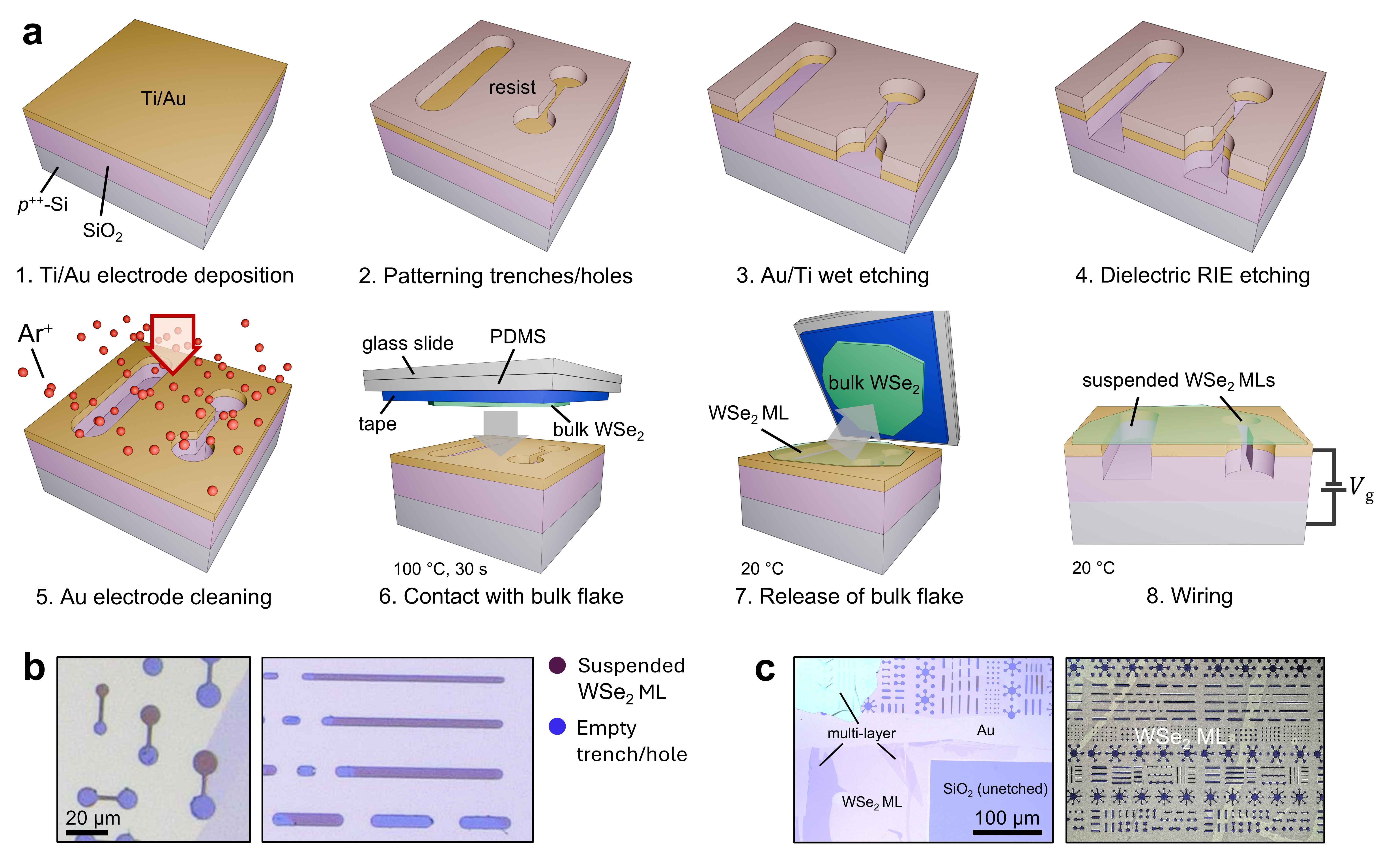}
  \caption{(a) Schematic illustration of the fabrication steps for gate-tunable suspended WSe$_2$ MLs. (b) Optical microscope images of suspended WSe$_2$ ML regions spanning large holes and trenches. (c) Optical microscope image showing the termination of the ML at the edge of the top contact electrode (left), and a contrast-enhanced image highlighting the lateral extent of the large-area WSe$_2$ monolayer, spanning several hundred micrometers (right).}
  \label{fig:figure1}
\end{figure*}

The fabrication workflow is summarized in Fig. \ref{fig:figure1}a. We begin with a heavily doped Si/SiO$_{2}$ (0.5 mm / 1026 nm) substrate on which a patterned Ti/Au (5 nm / 15 nm) electrode is deposited (step 1). This metal layer serves both as the top contact electrode and as the fresh Au surface required for GAE. Suspended regions are then defined by photolithography (step 2), followed by selective wet etching of Au/Ti (step 3). The SiO$_{2}$ is then partially etched via reactive-ion etching (RIE) using CHF$_{3}$ (step 4), leaving a residual SiO$_{2}$ thickness of $\sim$254 nm at the bottom of the etched regions, corresponding to a trench depth of $\sim$791 nm (including the Ti/Au thickness), such that the underlying Si substrate is not exposed. After resist removal and pre-cleaning with organic solvents, the patterned Au electrode is subjected to Ar-ion cleaning using a broad-beam ion source (step 5), which physically sputters the surface, removing adsorbates and a few atomic layers of Au (see Supporting Information). Ar-ion cleaning provides a contamination-free Au surface for GAE while preserving the electrical structure of the device, eliminating the need to re-evaporate a continuous Au film over the hole-patterned substrate, as in previous approaches \cite{huang2022efficient, geilen2025situ}. The approach is supported by established Ar plasma-cleaning studies for exfoliation of TMD MLs \cite{wu2024gold} and wafer bonding studies \cite{yamamoto2019comparison} demonstrating pristine Au surfaces. Immediately after Ar-ion cleaning, the sample is heated on a hot plate ($\approx$100 °C, 30 s) and a freshly cleaved WSe$_{2}$ bulk flake is gently pressed onto the Au surface (step 6), following protocols similar to those used in high-yield GAE \cite{heyl2023only}. After cooling to room temperature, the bulk crystal is slowly peeled away (step 7), leaving large ML regions that span across the etched holes and trenches. Then, the electrical contacts are wire-bonded to apply a gate voltage ($V_{\mathrm{g}}$) between the Au top contact electrode and the doped Si back gate (step 8). The gate voltage enables electrostatic control of doping and electromechanical strain, allowing systematic tuning of excitonic species \cite{hernandez2022strain}.

The resulting suspended structures are shown in Fig. \ref{fig:figure1}b. The color contrast of WSe$_{2}$ ML over cavities allows straightforward identification of suspended areas. Figure \ref{fig:figure1}c further shows the large-area, selective coverage of the ML on the patterned Au electrode, with a clear termination at the unetched SiO$_2$ boundary. The uniform coverage of the ML is enabled by the low roughness of the Au film under optimized GAE conditions \cite{velicky2018mechanism}, confirmed by atomic-force microscope images (see Supporting Information), and by the freshness of the Au surface required for gold-assisted exfoliation.

Our technique belongs to a class of hybrid approaches that use Au films for exfoliation while incorporating patterned non-contact regions, eliminating the need for post-processing Au etching that could otherwise damage the ML. Hybrid meshed Au/polymer templates have been developed \cite{li2021dry}, as well as contamination-minimized Au/SiO$_2$ mesh structures \cite{wu2024gold}; however, these approaches typically do not yield fully suspended membranes. Template-stripped hybrid GAE integration strategies \cite{satterthwaite2024van} further expand the versatility of these platforms for gating and device integration, but the ML remains supported on SiO$_2$. GAE followed by transfer has enabled suspended MLs \cite{mariani2024two}, though cryogenic linewidths are typically broadened. Consequently, our fabrication technique provides contamination-minimized suspension, large lateral uniformity, reproducibility, and applicable gate-tunability.

\begin{figure*}
  \centering
  \includegraphics[width=\linewidth]{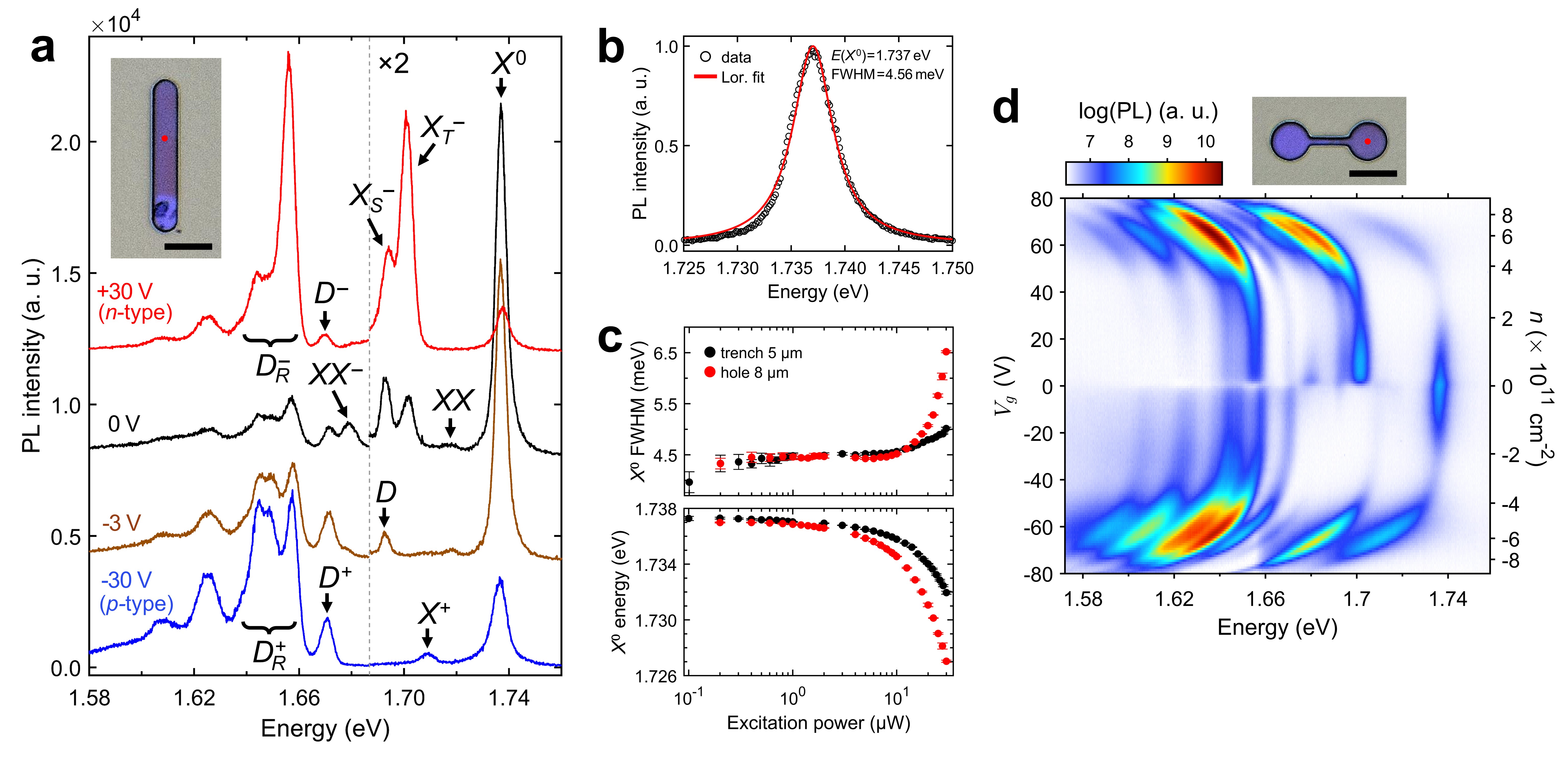}
  \caption{Low-temperature PL of suspended WSe$_2$ MLs. (a) PL spectra at 7 K under 532 nm continuous-wave (CW) excitation (1 $\mu$W) acquired on a 5 $\mu$m-wide trench (inset) at zero bias (0 V), near charge neutrality ($-3$ V), under $p$-doping ($-30$ V), and under $n$-doping ($+30$ V) conditions. Spectra are vertically offset for clarity. (b) Lorentzian fit of the normalized neutral-exciton peak at zero bias from (a). (c) Neutral exciton linewidth and energy as a function of excitation power, measured at the center of a 5 $\mu$m-wide trench and an 8 $\mu$m-diameter hole. (d) Gate-voltage-dependent PL spectra under 532 nm laser excitation at 1 $\mu$W for a ML suspended over an 8 $\mu$m-diameter hole (inset). Scale bars: 10 $\mu$m. The red dots at the centers of the optical microscope images in (a) and (d) indicate the measurement positions.}
  \label{fig:figure2}
\end{figure*}

The intrinsic optical quality of the suspended ML is evaluated using PL at $T \approx 7$ K under continuous-wave (CW) excitation ($\lambda = 532$ nm). Low-temperature measurements minimize thermal broadening and reduce exciton--phonon scattering, enabling accurate identification of excitonic resonances. Figure \ref{fig:figure2}a shows PL spectra acquired at the center of a 5 $\mu$m-wide trench for different $V_{\mathrm{g}}$. The exciton $X^0$ energy of the suspended ML at zero bias, measured at the center of etched holes or trenches, typically lies in the range $1.73$--$1.74$ eV across the device. Within an energy window of $\sim$80 meV below $X^{0}$, upon electrostatic $n$- or $p$-doping induced by $V_{\mathrm{g}}$, the spectra reveal multiple excitonic species: positive and negative trions ($X^+$ and $X^-$, including resolved triplet and singlet branches), dark exciton ($D$), charged dark excitons ($D^+$ and $D^-$), biexciton ($XX$), and charged biexciton ($XX^-$), consistent with the spectral fine structure reported for high-quality hBN-encapsulated WSe$_2$ \cite{courtade2017charged,robert2017fine, barbone2018charge, ye2018efficient, liu2019gate}. The observed PL spectra exhibit a well-separated set of narrow excitonic resonances, consistent with reduced extrinsic disorder and minimal spectral broadening in the suspended MLs. Focusing on resonances near $X^0$, for a small offset bias ($V_{\mathrm{g}}$ = $-$3 V) the $X^-$ resonance, which is still visible at zero bias, is suppressed, leaving only $D$ and $X^0$. The weak dark-exciton emission is consistent with previous reports showing that its out-of-plane dipole can be detected due to the finite numerical aperture of the objective lens collecting light at oblique angles \cite{wang2017plane}. At energies below $1.66$ eV, additional peaks attributed to the phonon replicas ($D^+_{R}$ and $D^-_{R}$) of charged dark excitons and defect-bound exciton complexes are observed in the spectra \cite{liu2019valley, li2020phonon, he2020valley, rivera2021intrinsic}. The relatively strong intensity of the phonon replicas is attributed to optical interference effects \cite{mariani2024two} and dipole-dependent emission \cite{liu2019valley}, which enhances the collection of phonon-assisted emission.

The neutral exciton $X^{0}$ exhibits a narrow Lorentzian PL lineshape (Fig. \ref{fig:figure2}b) at low excitation power, where laser-induced broadening is suppressed. Under weak CW excitation ($\sim$1 $\mu$W), $X^{0}$ exhibits linewidths in the range $4.3$--$4.6$~meV across the device. To quantify power-induced thermal effects, we measured the exciton linewidth and energy as a function of excitation power at $V_{\mathrm{g}} = 0$ V (Fig.~\ref{fig:figure2}c), at the center of both a 5 $\mu$m-wide trench and an 8 $\mu$m-diameter hole. As expected from the reduced thermal conductivity of suspended TMD membranes, in agreement with previous reports on suspended WSe$_2$ and MoSe$_2$ \cite{saleta2022unraveling}, the linewidth increases monotonically with excitation power, accompanied by an energy shift exceeding $1$ meV above $\sim$10 $\mu$W. Therefore, we performed the following measurements in the low-power regime (0.2--4 $\mu$W), where the linewidth remains power independent.

The high optical quality achieved by the transfer-free GAE-based fabrication enables us to resolve multiple closely spaced PL emission lines and to examine their simultaneous dependence on strain and electrostatic doping induced by the gate voltage. To clearly resolve strain-related effects within the experimentally accessible gate-voltage range, we focused a different suspended ML spanning a circular hole with a diameter of 8 $\mu\mathrm{m}$, which provides larger electromechanical deflection than the 5 $\mu$m-wide trench. The corresponding gate-dependent PL spectra are shown in Fig. \ref{fig:figure2}d. Electrostatic gating of suspended WSe$_2$ MLs, which simultaneously enables mechanical deflection and carrier-density modulation, has been established in earlier works \cite{hernandez2022strain, lee2023electric}. While our device operates on the same physical principles, the transfer-free GAE approach avoids polymer handling and yields reduced spectral broadening with more clearly resolved excitonic resonances, compared to what is typically observed in suspended monolayers fabricated by conventional polymer-transfer and multi-step assembly approaches \cite{morell2016high, wu2025modulation}. In Fig. \ref{fig:figure2}d, electrostatic doping gives rise to charged-exciton complexes and is accompanied by a pronounced redshift and intensity modulation of the excitonic resonances, including $X^0$, reflecting the combined influence of electrostatic doping \cite{chernikov2015electrical}, strain \cite{hernandez2022strain}, and optical interference \cite{mariani2024two}. The redshift is attributed to carrier-induced bandgap renormalization together with electrostatically induced strain arising from gate-voltage-driven membrane deflection in the suspended structure. The distinct contrast between spectra dominated by positively and negatively charged exciton species for small changes of $V_{\mathrm{g}}$ around zero bias indicates that the ML is close to charge neutrality, suggesting negligible residual doping. 

\begin{figure*}
  \centering
  \includegraphics[width=\linewidth]{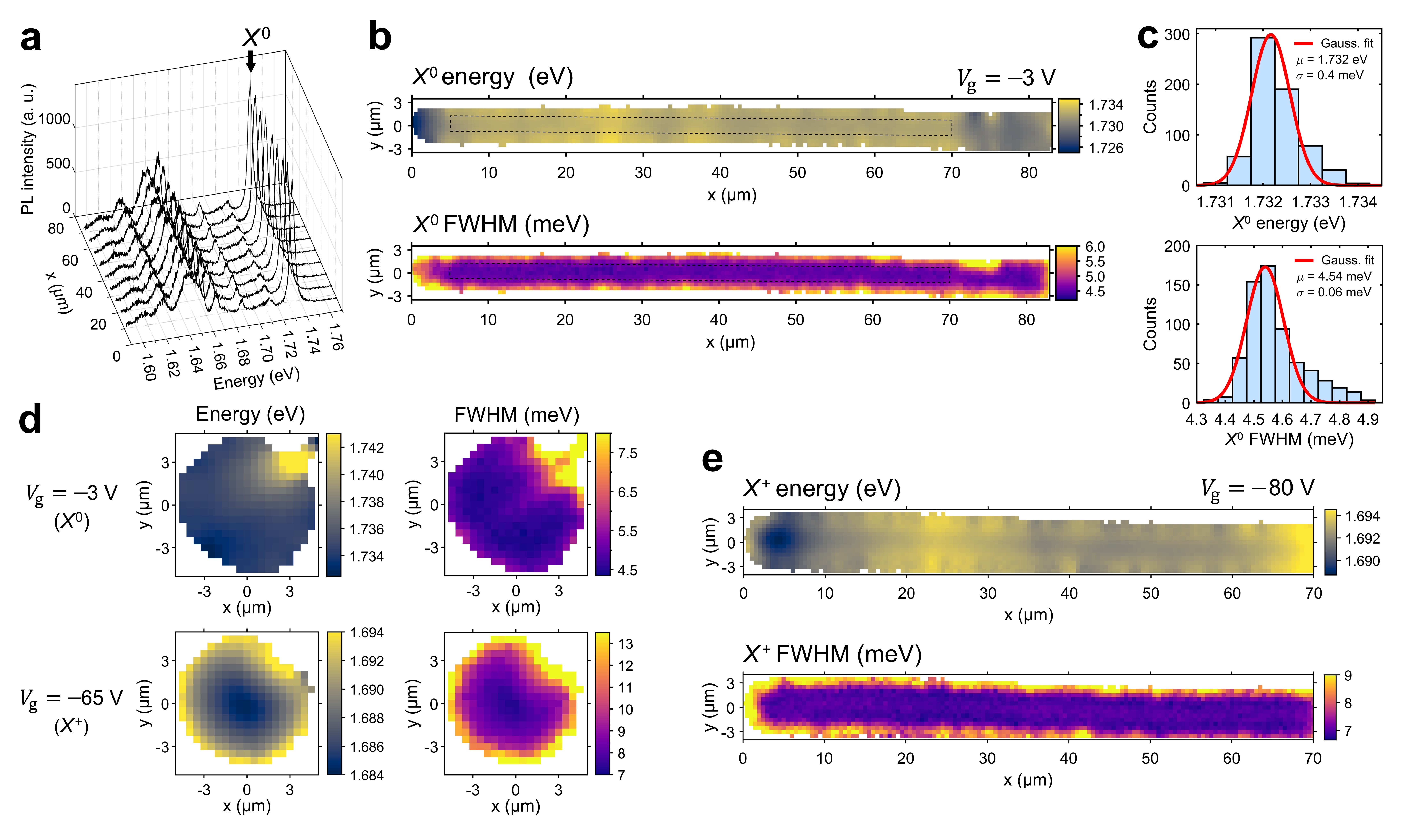}
  \caption{Spatial PL mapping in suspended WSe$_2$ MLs. (a) PL spectra acquired along the center line of a ML suspended over a long trench at $V_{\mathrm{g}} = -3$ V.  (b) Spatial PL maps of the $X^{0}$  energy and full width at half maximum (FWHM), extracted from Lorentzian fits of PL spectra for the trench in (a). (c) Histograms of the $X^{0}$  energy and FWHM extracted from the homogeneous region indicated by the dashed area in (b). (d) Spatial PL maps of the $X^{0}$ and $X^{+}$ energies and linewidths for a ML suspended over an 8 $\mu$m-diameter circular hole at $V_{\mathrm{g}} = -3$ V and $V_{\mathrm{g}} = -65$ V. (e) Spatial PL maps of the $X^{+}$ energy and linewidth for a ML suspended over a 4 $\mu$m-wide trench at $V_{\mathrm{g}} = -80$ V.}
  \label{fig:figure3}
\end{figure*}

A key advantage of our technique is the fabrication of extended suspended regions with excellent spatial homogeneity in the PL spectrum. The high optical quality of the ML allows us to reliably fit the exciton and trion peaks and to map their spatial dependence. Figure \ref{fig:figure3} shows the analysis of space-resolved PL spectra acquired at $T \approx 7$ K over trenches and holes to evaluate the exciton homogeneity in the largest exfoliated areas. 

The waterfall spectra in Fig. \ref{fig:figure3}a, taken along the center line of a $\sim$80 $\mu\mathrm{m}$-long and 3 $\mu$m-wide trench for $V_{\mathrm{g}}=-3$ V, exhibit nearly identical peak positions and lineshapes. Figure \ref{fig:figure3}b shows maps of the energy and full width at half maximum (FWHM) extracted from a Lorentzian fit of the $X^{0}$ peak. The mapping shows extended regions at the centerline of the trench where the $X^{0}$ energy is uniform around a mean value of $\mu \sim 1.732$~eV, with a mean FWHM of $\mu \sim 4.54$~meV. Figure \ref{fig:figure3}c shows a histogram of the statistics for the area of $2 \times 65~\mu\mathrm{m}^2$ enclosed by the rectangle highlighted in Fig. \ref{fig:figure3}b.  The standard deviation of the $X^{0}$ energy is $\sigma\sim$ 0.4 meV, whereas the $X^{0}$ FWHM exhibits a much smaller standard deviation of $\sigma\sim$ 0.06 meV, both well within the neutral-exciton linewidth, confirming high energy homogeneity along the suspended region. These statistics quantitatively demonstrate the excellent spatial homogeneity of the PL spectrum over a macroscopic suspended area. For comparison, spatially resolved statistical analyses in hBN-encapsulated TMD heterostructures report exciton-energy distributions with spreads of several meV \cite{wierzbowski2017direct}, with variability commonly attributed to residual strain and interfacial inhomogeneities \cite{rodek2021local}. In our sample, small variations (1–2 meV) in the potential landscape are observed across the suspended ML, likely caused by unwanted strain at the edges introduced during the exfoliation. We also attribute part of the gradients observed at the edges to the finite spot size of the excitation laser ($\sim$1.5 $\mu$m), which averages the PL signal with that of the ML placed over the Au film. 

Thus, our approach demonstrates spatial homogeneity over length scales of tens of micrometers, substantially exceeding the typically isolated spots of homogeneous regions accessible with polymer-based transfer and multi-step assembly methods \cite{castellanos2014deterministic, pizzocchero2016hot}, where limited flake sizes, local wrinkling, and interfacial residues can introduce pronounced spatial variations in the potential landscape \cite{raja2019dielectric}. While suspension has been shown to improve the spatial homogeneity of excitonic emission even in high-quality encapsulated MLs by relaxing local strain inhomogeneities \cite{zhou2020controlling}, the suspended regions in such devices are typically limited to only a few micrometers. In contrast, the transfer-free GAE process employed here yields millimeter-scale exfoliated MLs with high lateral continuity, which enables suspended regions extending over tens of micrometers while preserving optical quality and excitonic uniformity.

Applying a gate voltage to such homogeneous suspended MLs enables controlled access to excitonic species and further engineering of the potential landscape. Figures \ref{fig:figure3}d and \ref{fig:figure3}e further illustrate the spatial response of suspended MLs under $V_{\mathrm{g}}$ for an 8 $\mu$m-diameter hole and 4 $\mu$m-wide trench, respectively. Due to the suppression of $X^0$ at large $V_\mathrm{g}$, the spectrally isolated $X^{+}$ resonance is used to map the spatial response, as the $X^{-}$ emission exhibits a closely spaced doublet.

In the circular hole geometry (Fig. \ref{fig:figure3}d) at $V_\mathrm{g}=-3$ V, the $X^0$ energy typically shows a smooth gradient due to residual strain from exfoliation and the asymmetry introduced by the venting channel. The exciton linewidth far from the edges is typically $\sim$4.5 meV. Electrostatic gating induces both membrane deflection and charge accumulation. The gate-induced strain profile in circular holes preserves a radial symmetry, leading to concentric energy contours, similar to what is observed in gated suspended MoSe$_2$ MLs \cite{lee2023electric}.

In the trench geometry (Fig.~\ref{fig:figure3}e), spatial maps of $X^+$ at $V_{\mathrm{g}}=-80$ V show a largely uniform response across the suspended region. The corresponding $X^{+}$ FWHM map remains homogeneous, indicating that the optical quality is preserved under gating and that no pronounced spatially localized broadening is introduced across the suspended membrane.

This combination of geometric symmetry, controlled carrier injection, and spatial uniformity provides a suitable platform for transport-based and valley-dependent studies over extended suspended regions. Related strain-engineering approaches have demonstrated quasi-one-dimensional exciton channels and guided exciton diffusion in TMD MLs, for example using wrinkle architectures \cite{kim2024strained} or nanowire-induced strain in encapsulated MLs, highlighting the broader applicability of symmetry-defined strain landscapes for exciton transport \cite{dirnberger2021quasi}.

\begin{figure*}
  \centering
  \includegraphics[width=\linewidth]{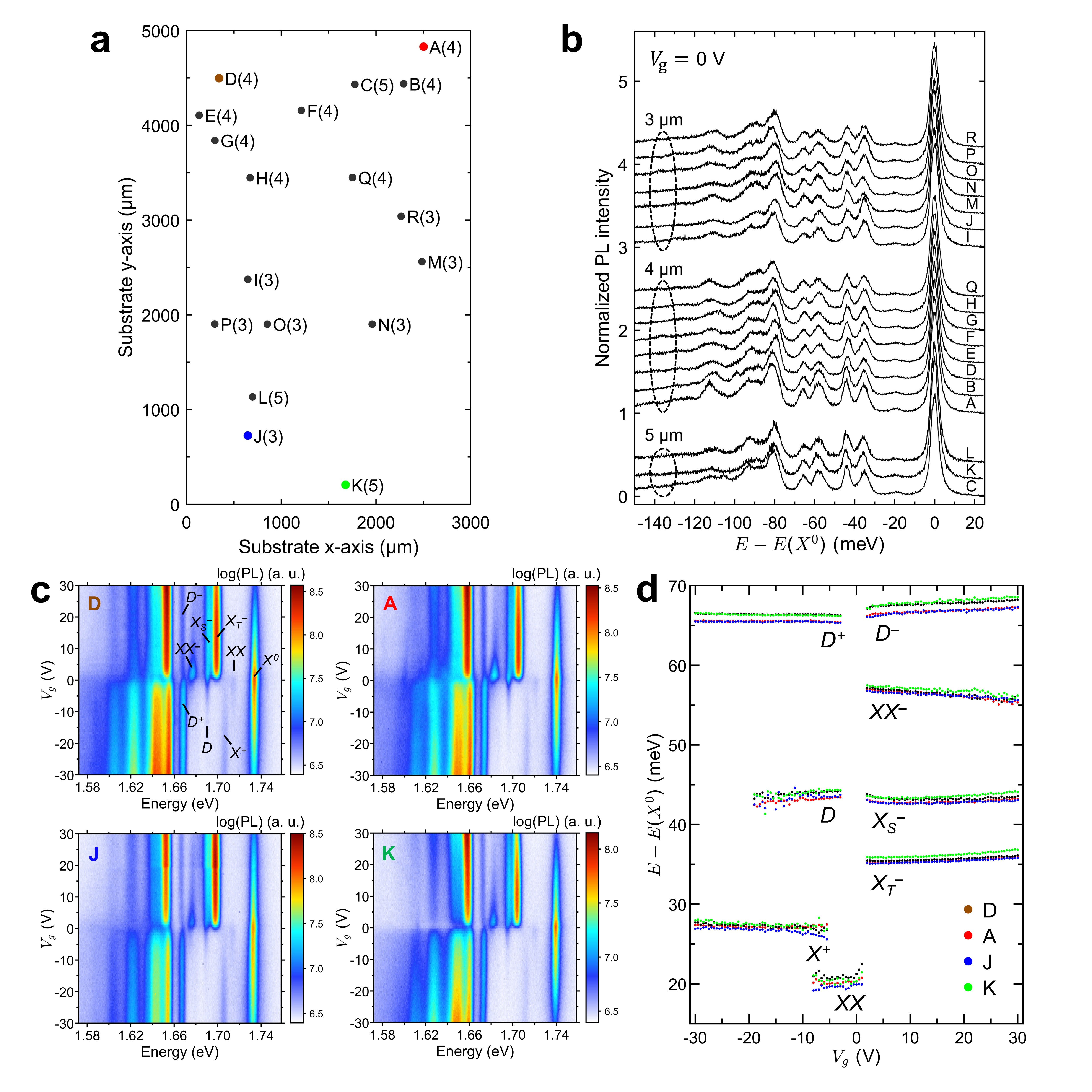}
  \caption{Spatial reproducibility of the PL across multiple suspended WSe$_2$ MLs. (a) Schematic map of the device showing the spatial positions of the analyzed suspended MLs spanning trenches with widths of 3, 4, and 5 $\mu$m; the number in parentheses indicates the trench width. (b) Normalized PL spectra acquired at the positions indicated in (a), plotted relative to the $X^{0}$ energy. Spectra are vertically offset for clarity. (c) Gate-dependent PL maps for four representative suspended MLs located near the corners of the device. (d) Relative energy shifts of the main excitonic resonances, plotted with respect to the $X^{0}$ energy, as a function of $V_{\mathrm{g}}$ for the four positions shown in (c). }
  \label{fig:figure4}
\end{figure*}

Beyond local uniformity, our fabrication yields suspended MLs whose optical response remains highly reproducible across millimeter-scale distances on a single device. Figure \ref{fig:figure4}a shows the absolute positions of the analyzed trenches with widths of 3, 4, and 5 $\mu$m located a few millimeters apart. For each suspended ML, we acquired PL spectra at zero bias at the center of the trenches to avoid edge-induced strain variations, and we plot the spectra shifted relative to the neutral exciton energy (Fig. \ref{fig:figure4}b). All these suspended regions originate from the same ML sheet, detached from a single bulk flake during one exfoliation step. Across all measured positions, the PL spectra exhibit the same set of excitonic resonances, with consistent relative energy separations and no evidence of localized defect-related emission. This reproducibility is further confirmed across independent fabrication batches (see Supporting Information). The absence of pronounced position-dependent spectral variations indicates that the observed PL response is dominated by excitonic emission intrinsic to the suspended WSe$_2$ ML rather than by defect-bound or environment-induced contributions. 

To further probe the reproducibility of the excitonic response of the spatially uniform suspended MLs under electrical tuning, we measured suspended MLs on four different trenches located near the corners of the top contact electrode (Fig.~\ref{fig:figure4}c). For simplicity, we applied modest gate voltages ($\left|V_\mathrm{g}\right|\leq30 $ V) to emphasize doping-induced spectral evolution while minimizing strain effects, which may depend on the different geometry of the suspended area. The excitonic species are consistent across the four measured locations, both in the number of resolved resonances and in their evolution. The relative energy spacing and ordering of excitonic resonances are consistent with previously reported spectra of WSe$_2$ encapsulated in hBN \cite{courtade2017charged, liu2019gate} and resonances could be assigned accordingly.  

Figure \ref{fig:figure4}d summarizes the energy shift of the excitonic resonances relative to the exciton energy as a function of $V_\mathrm{g}$ for each dataset in Fig. \ref{fig:figure4}c, obtained by fitting the spectra with multiple Lorentzian peaks. These reproducible emission lines enable reliable quantitative determination of the relative energies of excitonic species in suspended MLs. The relative binding energies were extracted at $V_{\mathrm{g}} = \pm 5~\mathrm{V}$ ($n \sim 1 \times 10^{11}~\mathrm{cm}^{-2}$), corresponding to weak electrostatic doping near charge neutrality. Referenced to $X^{0}$, we obtain $\Delta E_{X^{+}} = 27.0~\mathrm{meV}$, $\Delta E_{X^{-}_{\mathrm{T}}} = 35.5~\mathrm{meV}$, $\Delta E_{X^{-}_{\mathrm{S}}} = 42.9~\mathrm{meV}$, and $\Delta E_{D} = 43.2~\mathrm{meV}$, with a trion fine-structure splitting of $7.4~\mathrm{meV}$. Referenced to $D$, the charged dark excitons yield $\Delta E_{D^{+}} = 22.5~\mathrm{meV}$ and $\Delta E_{D^{-}} = 24.0~\mathrm{meV}$. The biexciton and charged biexciton have $\Delta E_{XX} = 21.0~\mathrm{meV}$ (from $X^{0}$) and $\Delta E_{XX^{-}} = 36.0~\mathrm{meV}$ (from $XX$). The close agreement among all measured suspended regions confirms that both the optical quality and the electrostatic response of the ML are highly reproducible across spatially separated locations on the same device.

Table \ref{tbl:binding} summarizes the binding energies of the excitonic resonances and compares them with previously reported values for MLs encapsulated in hBN. While being systematically higher by a few $\mathrm{meV}$ with respect to previous reports, these values are consistent with reduced dielectric screening in the suspended (air) geometry. This level of uniformity across spatially separated suspended regions is consistent with the intrinsic nature of the excitonic spectrum obtained through a contamination-free exfoliation technique.

\begin{table*}
  \caption{Relative binding energies (in meV) of excitonic resonances in suspended WSe$_2$ (this work), compared to representative values reported for hBN-encapsulated WSe$_2$ MLs from the literature. The values for the suspended MLs are extracted from Fig.~\ref{fig:figure4} at $V_{\mathrm{g}} = \pm 5$ V and are approximate.}
  \label{tbl:binding}
  \centering
  \begin{tabular}{lcccccccc}
    \hline
    & $X^{0}\!\rightarrow\!X^+$ & $X^{0}\!\rightarrow\!X^-_{\mathrm{T}}$ & $X^{0}\!\rightarrow\!X^-_{\mathrm{S}}$ & $X^{0}\!\rightarrow\!D$ & $D\!\rightarrow\!D^+$ & $D\!\rightarrow\!D^-$ & $X^{0}\!\rightarrow\!XX$ & $XX\!\rightarrow\!XX^-$ \\
    \hline
     \shortstack{Suspended\\(this work)}
      & 27.0 & 35.5 & 42.9 & 43.2 & 22.5 & 24.0 & 21.0 & 36.0\\
    \shortstack{hBN encap.\\(literature)}
      & 21\cite{courtade2017charged}
      & 29\cite{courtade2017charged}
      & 35\cite{courtade2017charged}
      & 40\cite{robert2017fine} 
      & 14\cite{liu2019gate} 
      & 16\cite{liu2019gate} 
      & 17-20\cite{barbone2018charge, ye2018efficient} 
      & 33\cite{barbone2018charge, ye2018efficient}  \\
    \hline
  \end{tabular}
\end{table*}

In summary, we demonstrated that suspended WSe$_2$ MLs exhibit a spatially uniform excitonic response over extended lateral dimensions. Spatially resolved cryogenic PL revealed reproducible neutral-exciton emission across suspended regions up to $\sim$80 $\mu$m, with linewidths as low as $\sim$4.5 meV and minimal spatial variation in exciton energy and linewidth, indicating strongly reduced inhomogeneous broadening. This high degree of exciton uniformity was enabled by a transfer-free GAE onto pre-patterned Au electrodes, which avoided polymer transfer while remaining compatible with electrostatic gating. Gate-dependent measurements further demonstrated the controlled evolution of multiple excitonic species across the suspended regions. Beyond enabling spatially homogeneous exciton spectroscopy, the realization of long, narrow suspended trenches provided a versatile device geometry for future experiments requiring extended quasi-one-dimensional channels. Together, these results established suspended WSe$_2$ MLs as a promising platform for investigating fundamental excitonic physics in a spatially uniform environment.
\section*{Acknowledgements}
The authors would like to thank Professor Hideki Gotoh for fruitful discussions.

\section*{Supporting information}
Supporting Information: Experimental procedures for sample preparation, including top contact electrode fabrication, heat treatments, etching, Ar-ion cleaning, transfer-free gold-assisted exfoliation, and device assembly. Additional details on the photoluminescence measurement setup and analyses of sample quality, optical response, and reproducibility are provided.

\bibliography{references_260406.bib}

\newpage


\clearpage


\section*{Supplementary material}

# Supporting Information

**Uniform Narrow Excitonic Spectrum in Large-Area Suspended WSe$_2$ Monolayers**

Giacomo Mariani*[1], Riccardo Lodo[1], Keigo Matsuyama[1], Yoji Kunihashi[1], Taro Wakamura[1], Satoshi Sasaki[1], Louis Smet[2], Makoto Kohda[2,3,4], Junsaku Nitta[1,2], and Haruki Sanada[1]

[1] NTT Basic Research Laboratories, NTT, Inc., Atsugi 243-0198, Japan

[2] Department of Materials Science, Graduate School of Engineering, Tohoku University, Sendai 980-8579, Japan

[3] Center for Science and Innovation in Spintronics, Tohoku University, Sendai 980-8577, Japan

[4] Quantum Materials and Applications Research Center, National Institutes for Quantum Science and Technology, Takasaki 370-1292, Japan

Corresponding author: giacomo.mariani@ntt.com

**S1. Sample preparation**

*Top Electrode Fabrication*

Heavily p$^+$-doped silicon substrates (0.001–0.005 Ω·cm) with a thermally grown SiO$_2$ layer of ~1026 nm thickness and an ultrasmooth surface roughness (RMS ~0.10 nm in a 2 × 2 µm$^2$ area) were used as the base for the gate-tunable devices. The Ti/Au films were deposited by electron-beam evaporation at rates of 0.5 Å/s. AFM images of the pristine Si-SiO$_2$ substrate and of the substrate after deposition of a Ti (5 nm)/Au (15 nm) metal stack are shown in Fig. S1. The surface roughness of the Au film, just after the deposition was found to be RMS~0.15 nm in a 2 × 2 µm$^2$ area. AFM images were plane-leveled prior to roughness analysis.

Since the surface roughness of the Au electrode critically affects the exfoliation quality of WSe$_2$, we tested two fabrication methods for the top contact electrode:

- Method 1 (used in main text): The Ti/Au film was first uniformly deposited on the SiO$_2$ surface after organic cleaning and a mild UV–ozone treatment. A photolithographic pattern was then defined, followed by wet etching of Au and Ti to shape the top contact electrode. This approach ensures a clean SiO$_2$/Ti interface and was used for all data presented in the main text.

- Method 2 (for integration into other device architectures): A conventional bilayer resist (LOR/S1813) was spin-coated before photolithography. After development, a

brief UV–ozone treatment (1 min at room temperature) was applied to remove resist residues in the developed area, followed by Ti/Au deposition and lift-off. We did not find significant differences between method 1 and this method in terms of exfoliation yield of the monolayer (ML) after Ar-ion cleaning.

*Considerations about heat treatments during fabrication*

The temperatures used through the fabrication of our devices shown in the main text did not exceed 110 °C, since we noted that bake temperatures exceeding 160 °C increased the surface roughness of the Au film, as shown in Fig. S2. In the case of Method 2, where a bi-layer resist was used, the baking temperature of the resist was kept at a maximum of 150 °C.

*Etching of the Suspended Holes and Trenches*

A positive photoresist (MICROPOSIT™ S1813™) was spin-coated and baked for 3 min at 110 °C. After photolithography, Au and Ti were selectively removed by wet etching (iodine-based solution for Au, hydrogen peroxide-free acid solution for Ti). The $SiO_2$ layer was then partially etched by reactive-ion etching (RIE) using $CHF_3$ gas to form holes and trenches. The remaining $SiO_2$ thickness at the bottom of the etched regions was approximately 254 nm. The resist was removed by immersion in N-methyl-2-pyrrolidone (NMP) heated at 60 °C.

*Argon-Ion Cleaning*

Prior to Ar-ion cleaning, the device was cleaned sequentially in acetone, isopropanol, and deionized water. A mild UV–ozone treatment (1 min) was applied, followed by a hot plate treatment (100 °C, 1 min) to remove resist and organic residues that would otherwise cause surface unevenness after Ar sputtering. We confirmed that UV–ozone exposure up to 5 min did not modify the Au surface roughness. We did not use prolonged exposure to UV–ozone which has been reported to form a thin $Au_2O_3$ layer on the Au surface [S1]. The sample was then promptly inserted into vacuum for Ar-ion cleaning. For the devices presented in the main text, Ar ions were accelerated at 3 kV for 30 s at a pressure of $4.0 \times 10^{-3}$ Pa, etching approximately ~1 nm of Au (Fig. S3a). After the cleaning step, the device was held at $2 \times 10^{-5}$ Pa for 30 minutes to cool before venting.

A summary of the test with varying etching time is shown in Fig. S3b. To compare the surface roughness and etching depth after the Ar-ion cleaning, a transmission electron microscope (TEM) grid was used as a shadow mask. The Au film etched for 30 s exhibited a negligible change in the surface roughness (RMS~0.15 nm and ~0.17 nm before and after the etching, respectively). This nanoscale smoothness is essential for achieving the adhesion strength

and exfoliation uniformity characteristic of high-quality gold-assisted exfoliation. At longer etching times the surface roughness started to increase.

*Exfoliation of the monolayer and device assembly*

Within 1 minute after venting, the device was placed on a hot plate at 100 °C for 30 s. A freshly cleaved flux-grown $WSe_2$ bulk crystal (2Dsemiconductors™, Premium Ultra-Flat $WSe_2$ Synthetic Crystals) on blue adhesive tape was pressed onto the device for 30 s using a PDMS gel stamp (Gel-Pak®, X8 type) attached to a glass slide to apply uniform pressure. After cooling to room temperature, the tape was slowly peeled away, leaving the $WSe_2$ ML adhered to the Au surface and suspended over the etched cavities. Within a few minutes, the sample was transferred to a vacuum storage container (≈$10^2$ Pa) for transport to the measurement setup. The sample remained in the container for approximately 15-20 min. The device was mechanically anchored to a gold plate using vacuum grease and wire bonded to an IC chip for gate-voltage application. The total air exposure during these steps was approximately 30 min. Finally, the device was placed under high vacuum ($2 \times 10^{-5}$ Pa) in a helium-flow cryostat and cooled to 6.5–7 K for photoluminescence measurements. Because the exfoliation and suspension process yields only a limited number of intact suspended MLs, electrostatic gating can further act as a mechanical "selection" step: membrane deflection under bias causes structurally weaker suspended regions to collapse, while mechanically stable membranes remain suitable for optical characterization.

## S2. Photoluminescence measurements

In our devices, PL measurements were performed using a continuous-wave (CW) 532 nm laser with excitation power in the range 0.1–1 µW. The beam was focused onto the sample using an achromatic objective lens (MITUTOYO™, 50x LCD Plan APO NIR, NA = 0.42) positioned outside the cryostat, resulting in a spot diameter of approximately 1.5 µm. Emission spectra were recorded using a Princeton Instruments SP2500i spectrometer equipped with a 600 g/mm grating.

## S3. Considerations about the final quality of the suspended monolayer

Previously reported suspended MLs fabricated by gold-assisted exfoliation showed broadened bound exciton emission [S2] compared to our work. We cannot identify a single dominant mechanism responsible for the narrow linewidths observed in our devices; however, we observed gradual improvements associated with the following factors:

- Using the latest flux-grown $WSe_2$ bulk flake commercially available. The final quality of the suspended ML is strongly correlated with the quality of the bulk flake. Flux grown TMD flakes were reported to show the lowest density of defects [S3].

- Shortening the air exposure of the device to less than one hour after the exfoliation and inserting the sample into high vacuum as soon as possible, to limit airborne adsorbates.
- Avoiding baking the device in air at temperatures around 100-150 °C typically required for wire bonding with Au wires.
- Ensuring proper radiation shielding in the cryostat and avoiding elements at high temperature near the sample during measurements at low temperature.

We performed control measurements to assess potential changes in the PL linewidth of suspended $WSe_2$ MLs after air exposure up to 24 hours at room temperature. After progressive exposure to air, the sample was cooled at 7 K and PL map was acquired over a ML suspended over a 10 µm-diameter hole. The comparison of the PL maps of the exciton linewidth shows little difference for different exposures (Fig. S5). We could not evaluate a change in the quality on a time scale of minutes due to the unavoidable exposure to air during sample fabrication, transport, and device preparation. Previously, a rapid increase in wettability from airborne hydrocarbons and organic compound was shown to occur within 15-20 min from exposure to air [S4], which may cause a change in the final quality.

**S4. Structure of the device**

Evaluation of the linewidth at zero gate voltage may be affected by spectral narrowing or broadening originating from the Purcell effect. Following the procedure in [S5] and using the transfer matrix method, we calculated the reflection coefficient as a function of geometric parameters (Fig. S5a), from which we derived the normalized electric field intensity (Fig. S5b) and the exciton linewidth (Fig. S5c) at the ML interface. The quantities are calculated for an 8 µm-diameter hole, at an exciton energy of 1.735 eV+0.100 eV/(strain %), which takes into account the exciton shift due to strain induced by deflection. For the structure used in the main text, near zero deflection, the exciton linewidth is expected to be close to its intrinsic value.

**Supplemental References**

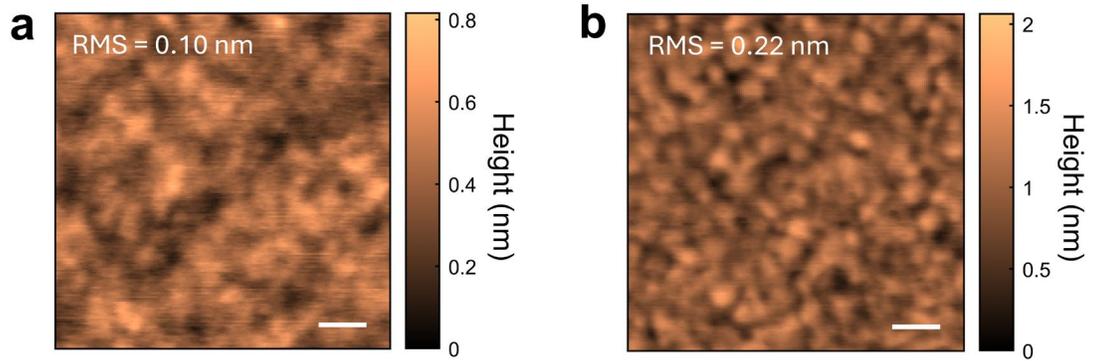

**Supplementary Figure 1** AFM images of (a) pristine Si/SiO$_2$ substrates and (b) with an electron-beam–evaporated Ti (5 nm)/Au (15 nm) metal stack. Scale bars: 100 nm.

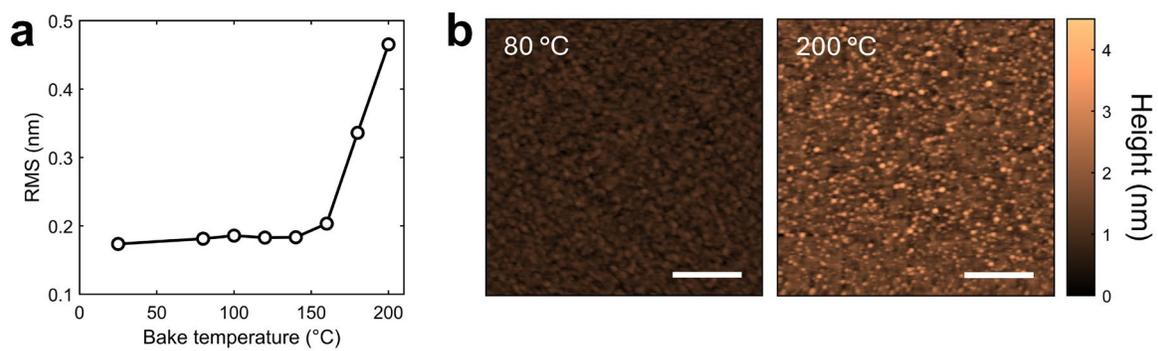

**Supplementary Figure 2** (a) Surface roughness of the Au film after 2 min hot plate baking at different temperatures. (b) AFM images of the Au surface after baking at 100 °C and 200 °C. Scale bars: 500 nm.

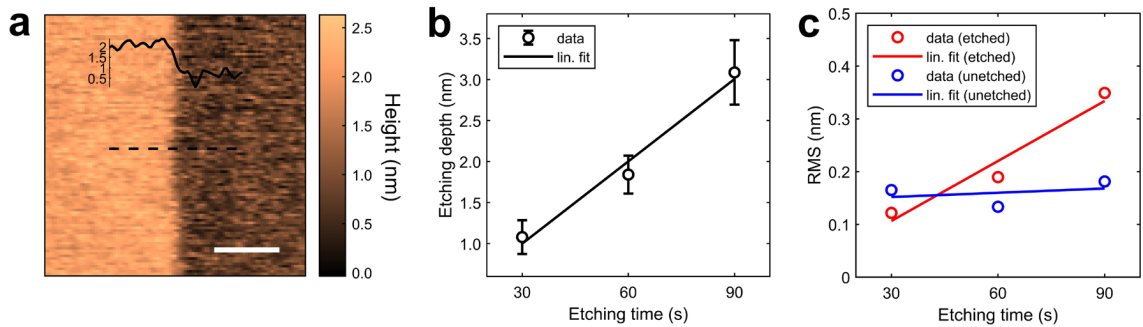

**Supplementary Figure 3** (a) AFM map of the Au surface etched by Ar ions at 3 kV for 30 s. A TEM was used as shadow mask to evaluate the etching depth. Scale bar: 500 nm. (b) Summary of the etching depths for 30 s, 60 s, and 90 s etching times. (c) Surface roughness of the Au film in the etched and masked areas for 30 s, 60 s, and 90 s etching times.

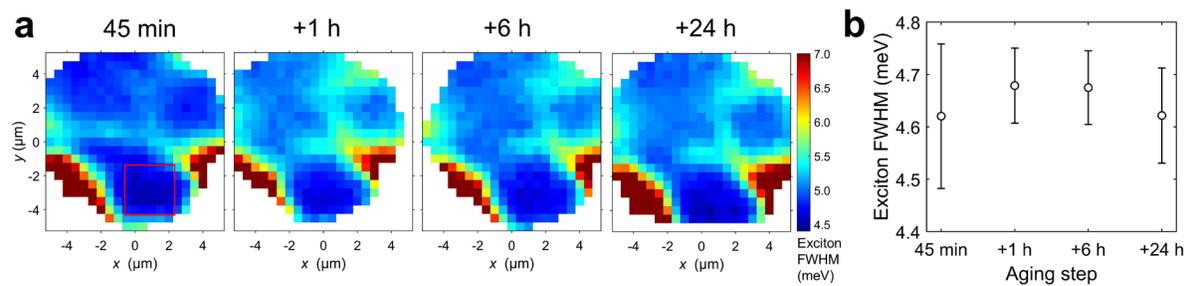

**Supplementary Figure 4** (a) Spatial maps of the exciton linewidth of a ML suspended over a 10 µm-diameter hole for increasing exposure to air. Excitation power ~4 µW. (b) Summary of the average FWHM calculated from a 2 × 2 µm$^2$ area highlighted by the red rectangle in (a).

9

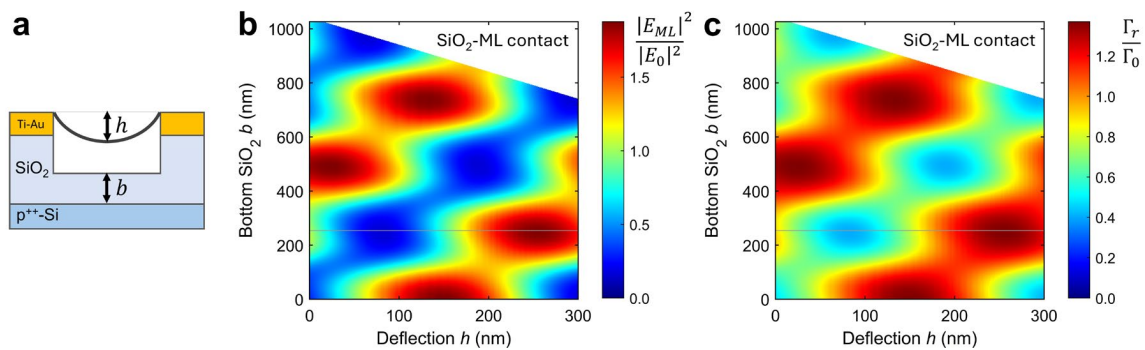

**Supplementary Figure 5** (a) Schematic representation of the device structure. (b) Calculated normalized electric field intensity and (c) exciton linewidth at the maximum deflection point of the ML, as a function of the ML deflection $h$ and residual $SiO_2$ thickness $b$. The grey line indicates the experimentally measured residual $SiO_2$ thickness (~254 nm) remaining at the bottom of the etched regions after RIE.



\end{document}